# Topological phonons in an inhomogeneously strained silicon-6: Possible evidence of the high temperature spin superfluidity and the second sound of topological phonons

Anand Katailiha[1‡], Paul C. Lou[1‡], Ravindra G. Bhardwaj[1], Ward Beyermann[2], and Sandeep Kumar[1,*]

[1] Department of Mechanical Engineering, University of California, Riverside, CA 92521, USA

[2] Department of Physics and Astronomy, University of California, Riverside, CA 92521, USA

[*] Corresponding author

[‡] Equal contribution

Email: sandeep.suk191@gmail.com

**Abstract**

The superposition of topological phonons and flexoelectronic charge separation in an inhomogeneously strain Si give rise to topological electronic magnetism of phonons. The topological electronic magnetism of phonons is also expected to give rise to stationary spin current or spin superfluidity. In this experimental study, we present possible evidence of spin superfluidity in an inhomogeneously strained p-Si thin films samples. The spin superfluidity is uncovered using non-local resistance measurement. A resonance behavior is observed in a non-local resistance measurement at 10 kHz and between 270 K and 281.55 K, which is attributed to the second sound. The observation of second sound and spatially varying non-local resistance phase are the evidences for spin superfluidity. The spatially varying non-local resistance with opposite phase are also observed in Pt/MgO/p-Si sample. The overall non-local responses can be treated as a standing waveform from temporal magnetic moments of the topological phonons.

Recently, the dissipationless current was reported in the freestanding MgO/Si bilayer structure[1]. However, the mechanistic origin of the dissipationless spin current was not well understood. In an MgO/p-Si sample, we hypothesized that the oxide layer deposited on top of Si thin film lead to the inhomogeneous strain. The oxide layer also leads to charge accumulation at the interface in addition to interfacial flexoelectronic effect[2] as shown in Figure 1 (a). It is noted that second order non-linearity[3] and optical second harmonic[4] has been reported in lightly doped Si waveguides due to inhomogeneous strain and interfaces. Whereas the flexoelectronic charge separation and flexoelectronic effect is expected to arise in highly doped Si thin film structure as shown in Figure 1 (a). The inhomogeneous strain give rise to spatially varying phonon frequency and dispersion, which is expected to give rise to topological phonons as shown in Figure 1 (a). The superposition of topological phonons and flexoelectronic charge separation give rise to topological electronic magnetism of phonon or topological dynamical multiferroicity as shown in Figure 1 (a). The dynamical multiferroicity[5] in the Si can be described as:

$$\boldsymbol{M}_t \propto \boldsymbol{P}_{FE} \times \partial_t \boldsymbol{P} \qquad (1)$$

where $\boldsymbol{M}_t$, $\boldsymbol{P}_{FE}$ and $\boldsymbol{P}$ are temporal magnetic moment, flexoelectronic effect and time dependent polarization of optical phonons, respectively. The dynamical multiferroicity can also be understood in the framework of topological electronic magnetism[6] of phonons described as:

$$\boldsymbol{M}_t \propto \frac{\partial n}{\partial z} \times f(\boldsymbol{A}, \boldsymbol{p}) \qquad (2)$$

where $P_{FE} \propto \frac{\partial \epsilon_{xx}}{\partial z} \propto \frac{\partial n}{\partial z}$ is flexoelectronic effect due to strain gradient $\left(\frac{\partial \epsilon_{xx}}{\partial z}\right)$ that give rise to gradient of charge carrier concentration $\left(\frac{\partial n}{\partial z}\right)$ and $\partial_t P \propto f(A, p)$ is time evolution of topological phonon polarization that is a function of Berry gauge potential $A$ and momentum $p$. The dissipationless spin current reported in MgO/Si thin films structures may have arisen from topological electronic magnetism of phonons.

In the part 1 of this series, we demonstrated a long-distance spin transport using transverse spin-Nernst effect measurement. The long-distance spin transport was expected to arise due to topological phonons in an inhomogeneous medium. In addition, we have also demonstrated spin-momentum locking and topological electronic magnetism of the phonons in part 2 and 3, respectively. The topological electronic magnetism of the phonons is expected to give rise to spatially varying spin distribution due to spin dependent electron-phonon scattering, which was the underlying cause of spin-Hall effect[5,7] reported previously. As a consequence, topological electronic magnetism of phonons can also give rise to a stationary spin current and spin superfluidity. The spin superfluidity can be defined as transport of spin angular momentum without significant dissipation[8,9]. The spin superfluidity has been topic of intense research since it can lead to efficient spintronics device applications. This motivated us to undertake experimental measurement to uncover possible evidence of spin superfluidity in inhomogeneously strained Si thin films. In this study, we present experimental evidence of spin superfluidity in an inhomogeneously strained highly doped Si thin films samples. The spin superfluidity was discovered using non-local resistance measurement as a function of temperature. The spin superfluidity was verified by experimental observation of second sound in non-local resistance measurement.

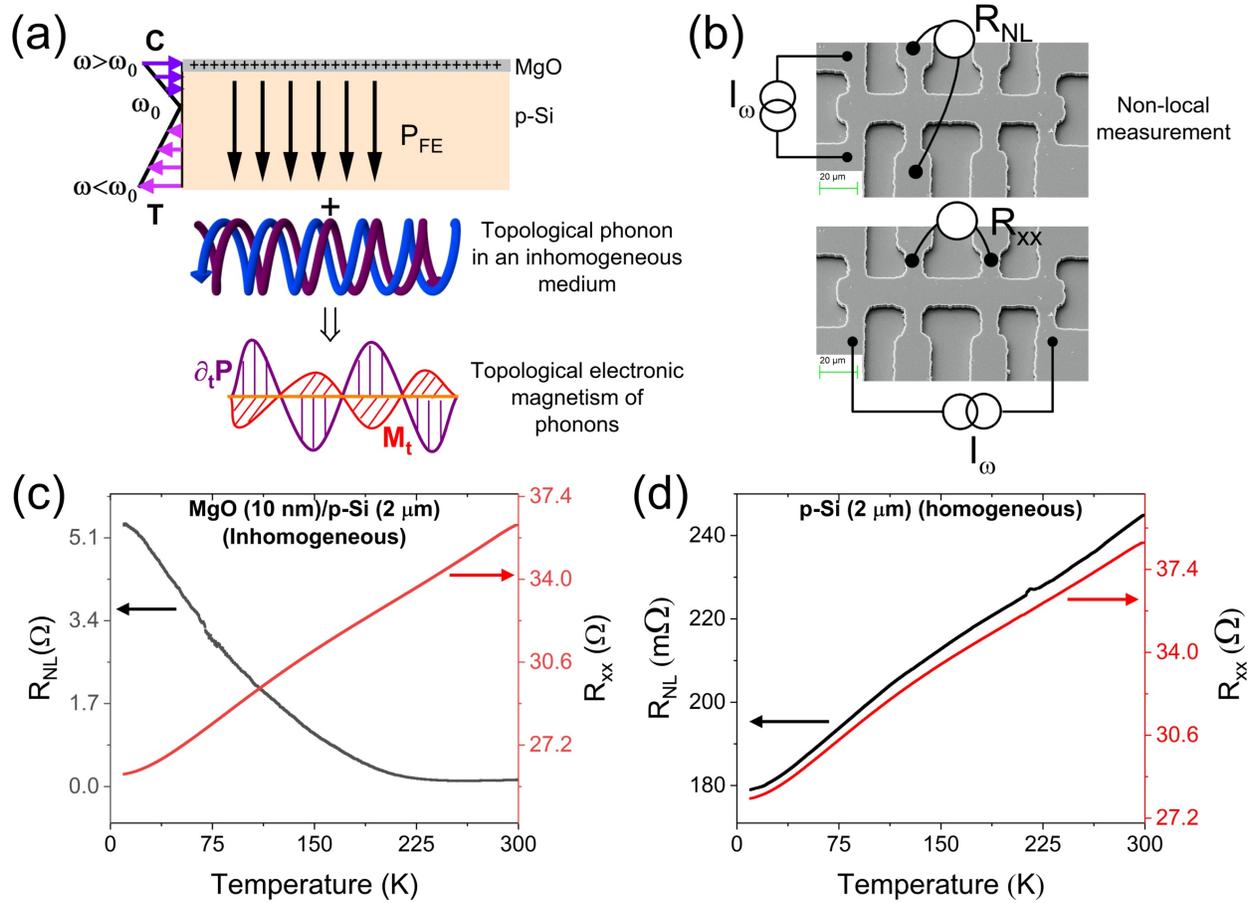

Figure 1. (a) A schematic showing the superposition of flexoelectric charge separation and topological phonons giving rise to topological electronic magnetism of phonons, (b) a representative scanning electron micrograph showing the experimental scheme for longitudinal resistance and non-local measurement. The non-local resistance and longitudinal resistance as a function of temperature from 300 K to 10 K for (c) MgO (10 nm)/p-Si (2 μm) (inhomogeneous) and (d) p-Si (2 μm) homogeneous samples.

We fabricated two p-Si (2 μm) samples in order to demonstrate the spin superfluidity. The first sample had 10 nm of MgO sputter deposited on top of the p-Si. We used Ar milling for 15 mins to etch the native oxide, which might have increased the charge accumulation at the interface. The 10 nm of MgO deposition would induce inhomogeneous strain in the sample as well as large flexoelectronic charge separation in the p-Si layer, which was expected to give rise to topological electronic magnetism of phonons. The second sample was a control sample with homogeneous p-Si. We did not make these samples freestanding as we had done in previous studies. The spin superfluidity is usually studied using non-local measurement technique. The non-local resistance can be described using van der Pauw's theorem as:

$$R_{NL} = R_{sq} e^{\frac{-\pi L}{w}} \qquad (3)$$

where $R_{sq} = \frac{\rho}{t}$, $\rho$, $t$, $L$ and $w$ are resistivity, thickness, length and width of channel. Hence, the non-local resistance measured as a function of temperature should exhibit behavior similar to longitudinal resistance if the non-local response was due to leakage current only. First, we measured the longitudinal resistance for both the samples using the experimental setup as shown in Figure 1 (b). We, then, measured the non-local resistance by passing the current across the junctions J1 and measuring the voltage drop across junction J2 using lock-in technique. The resulting responses are shown in Figure 1 (c,d) for MgO/p-Si and control p-Si sample, respectively.

In the MgO/p-Si sample, the magnitude of the non-local resistance at the room temperature (300 K) was similar to the estimates using van der Pauw's theorem, which indicated absence of any spin dependent response. However, the non-local resistance response increased as the temperature was reduced to 10 K as shown in Figure 1 (c). The non-local resistance at 10 K was ~20% of the longitudinal resistance and an order of magnitude larger than that due to leakage current. The behavior was opposite to the longitudinal resistance response and could not arise due to leakage current. This difference in response was attributed to the possible spin superfluidity from topological phonons in an inhomogeneous medium. It was noted that the non-local resistance was not observed in the measurement at junction J4 in this sample, which was 100 µm away from the source. Whereas the previous measurement on the freestanding sample showed the non-local resistance even at J4. We attributed this behavior to absence of heat transport in our current sample on substrate since it was not freestanding.

In case of the control sample, the magnitude of the non-local resistance was similar to estimates using van der Pauw's theorem. The non-local resistance behavior as a function of temperature was same as longitudinal resistance as shown in Figure 1 (d) expectedly. This suggested that the non-local resistance was due to leakage current and no spin current exists in the homogeneous p-Si sample. This study clearly showed that there was an additional response in case of inhomogeneous p-Si sample (MgO/p-Si). While we attributed the additional response to the spin superfluidity in case of inhomogeneous sample but there could be additional mechanisms that could potentially give rise to the observed behavior. The observed behavior was not expected to arise due to spin-Hall effect, which disappears at low temperatures in Si[5].

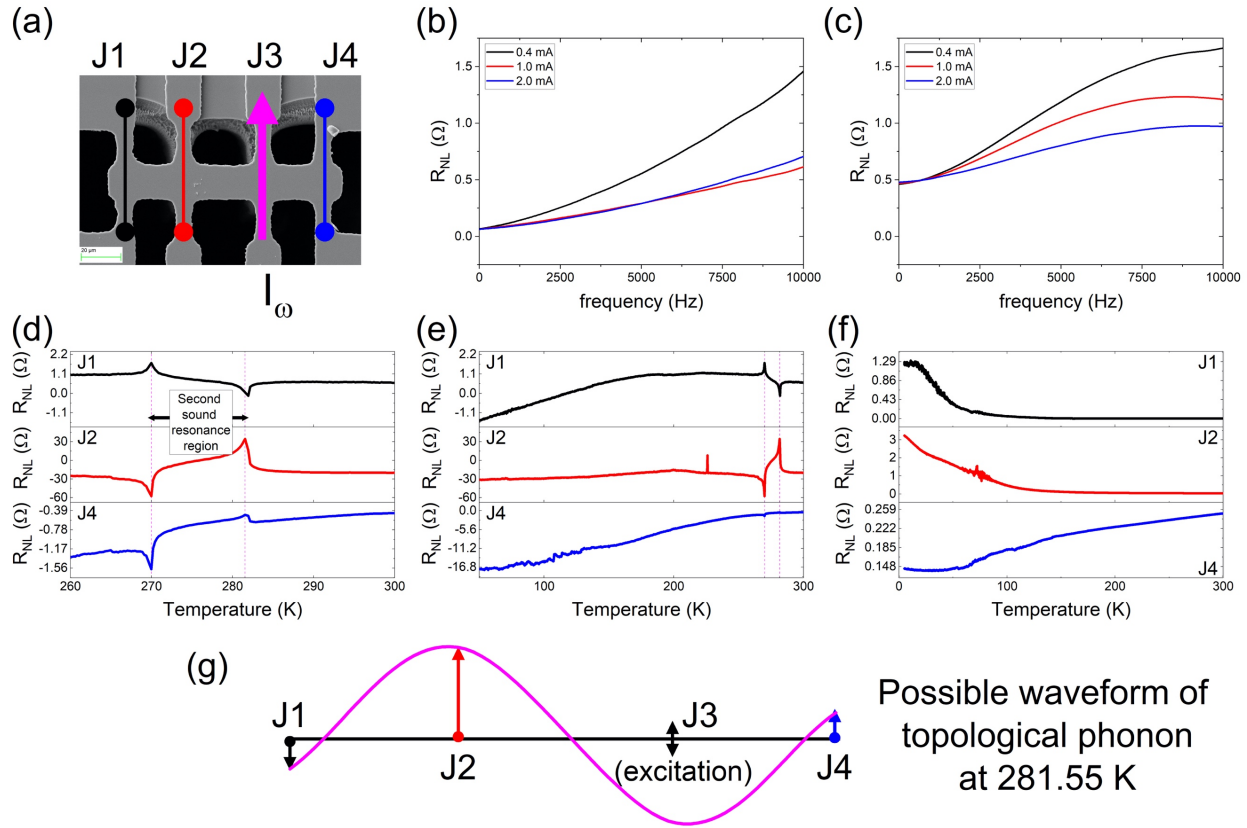

Figure 2. (a) a representative scanning electron micrograph showing the measurement scheme for non-local resistance across junctions J1, J2 and J4 while the current is applied across junction J3. The non-local resistance in MgO (2 nm)/p-Si (2 µm) freestanding sample as a function of frequency at (b) junction J2 and (c) junction J4 for an applied ac bias of 0.4 mA, 1 mA and 2 mA across junction J3. The non-local resistance measurement across junctions J1, J2 and J4 for 2 mA 10 kHz ac bias applied across J3 as a function of temperature (d) from 300 K to 260 K showing the resonance from second sound of topological phonons and (e) the complete response from 300 K to 50 K. (f) The non-local resistance measurement across junctions J1, J2 and J4 for 2 mA 37 Hz ac bias applied across J3 as a function of temperature from 300 K to 5 K. (g) A schematic showing the possible waveform of topological phonon corresponding to the resonance shown in (d-e).

We hypothesized that if the response was indeed due to spin superfluidity from topological phonons then we should observe the phenomenon called "second sound". In the superfluidity, the transport behavior is wave mediated and not diffusion driven. As a consequence, the second sound can be treated as a signature of superfluid behavior. In the non-local measurement scheme, a frequency dependent response can be used to characterize the spin superfluid and second sound behavior. In the next experiment, we chose a freestanding p-Si (2 µm) sample with 2 nm of MgO on top to achieve the in-plane thermal transport behavior. We applied a known current (0.4 mA, 1 mA and 2 mA) across the junction J3 and measured the non-local resistances across junctions J2 (40 µm) and J4 (30 µm) as shown in Figure 2 (a). The non-local resistance increased as a function of frequency as shown in Figure 2 (b,c). The non-local resistance measured at J4 showed a saturation behavior as the frequency was raised to 10 kHz as shown in Figure 2 (c).

This led us to choose current bias frequency of 10 kHz for temperature dependent response measurement. We applied a 2 mA 10 kHz current bias across the junction J3 and measured the non-local responses across junctions J1, J2 and J4 as shown in Figure 2 (a). The measured responses as a function of temperature from 300 K to 50 K are shown in Figure 2 (d-e) for J1 and J2 and J4 junctions. We observed that the responses showed a sharp change in amplitude at ~281.55 K and 270 K as shown in Figure 2 (d). At 281.55 K, the non-local resistances at J2 and J4 showed a peak while at J1 a valley was observed. The inverse was observed at 270 K. This sharp change in non-local resistance was similar to a frequency response function in a vibrating system. In our case, this behavior was attributed to the possible resonance from the advent of the second sound. Recently, Beardo et al.[10] reported the observation of the second sound in Ge using rapidly varying temperature field. They observed a sharp change in phase, which was attributed to the second sound. In our experiment, we use high frequency ac bias to rapidly change the temperature field. As a consequence, the wave transport lead to sharp change in the non-local resistance response observed in our measurement. We, then, measured the non-local response at 2 mA 37 Hz of ac bias applied at J3 junction as shown in Figure 2 (f). We did not observe any resonance behavior in this measurement. In the measurement at 10 kHz, the responses at J2 and J1 were out of phase with each other whereas they are in-phase at 37 Hz as shown in Figure 2 (e) and (f). In addition, the non-local responses measured at J2 and J4 were in-phase at 10 kHz even though they were on the opposite side of junction J3 spatially. This observed behavior was attributed to the topological phonons and expected waveform is shown in Figure 2 (g). This waveform was also consistent with the previously reported spin density wave in the part 5 of this series where wavelength of the spin density wave was expected to be ~140 µm. This measurement also demonstrated that the spin superfluidity was due to stationery spin current from topological phonons.

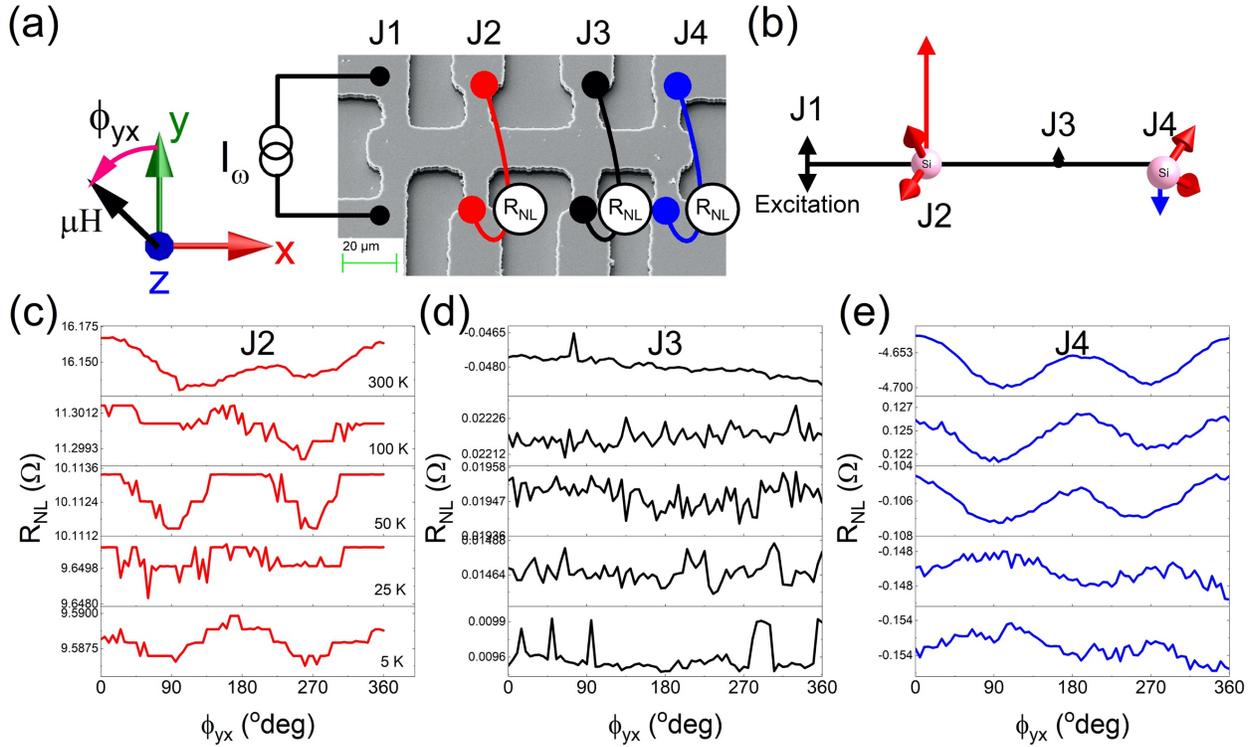

Figure 3. (a) A representative scanning electron micrograph showing the experimental scheme for angle dependent non-local resistance measurement where 2 mA 37 Hz ac bias w is applied across junction J1 and responses are measured across junctions J2, J3 and J4. (b) A schematic showing the spatially varying response from temporal magnetic moment. The angle dependent non-local resistance at an applied magnetic field of 4 T in the yx-plane measured at (c) J2, (d) J3 and (e) J4 junctions and at 300 K, 100 K, 50 K, 25 K and 5 K.

So far, we have measured the non-local response in an inhomogeneous p-Si sample where inhomogeneity was induced using MgO layer deposition. To present additional proof of wave mediated transport, we measured the angle dependent non-local resistance in the Pt (15 nm)/MgO/p-Si sample, where we have already demonstrated the dynamical multiferroicity[5]. We applied a 2 mA of current bias across the junction J1 as shown in Figure 3 (a). We, then, measured the angle dependent non-local resistance across junctions J2, J3 and J4 for an applied magnetic field of 4 T as shown in Figure 3 (a). The sample was rotated in the yx-plane as shown in Figure 3 (a) and measurement was done at 300 K, 100 K, 50 K, 25 K and 5 K. The measurement data for J2, J3 and J4 is shown in Figure 3 (c-e). The non-local resistance at J2 was larger than even the longitudinal resistance of the sample at 300 K. In addition, the symmetry of the angle dependent response was estimated to be using three signals having $\cos\phi_{yx}$, $\sin\phi_{yx}$ and $\sin^2\phi_{yx}$ symmetries. The angle dependent behavior was expected to arise due to angular relationship between magnetic field, momentum (**p**) and temporal magnetic moment ($M_t$). Here, the $\cos\phi_{yx}$ response was due to magnetic field being parallel and antiparallel to the component of the temporal magnetic moment of topological phonons. The $\sin\phi_{yx}$ and $\sin^2\phi_{yx}$ responses were due to magnetic field being parallel and antiparallel to the momentum of the topological phonons. The response at J3 was found to be insignificant and no angular symmetry was observed. In contrast, the response measured at junction J4 was much larger than that at J3 even though J4 was farther away from the current source. It also showed symmetry behavior similar to the measurement at J2 except with opposite phase. The non-local response decreased as a function of temperature as shown in Figure 3 (c-e). The response at J3 was always insignificant without any angle dependent behavior across all temperatures as shown in Figure 3 (d). Whereas the response at J4 was opposite to that of J2 except at 100 K. The angle dependent behavior was more pronounced at J4 as compared to J2 in spite of J2 being nearer to the current source. We have already demonstrated that the temporal magnetic moment in the Si were aligned along the <111> directions in the (110) cross-section plane. However, the longitudinal resistance demonstrated averaged contributions from each <111> direction whereas local topological phonon could be aligned along only one of the directions. Hence, the topological phonons having spin aligned along the <111> directions were expected to give rise to spin superfluidity mediated non-local response. For example- the response at J2 might arise from $M_t^{[1\bar{1}1]}$ and $M_t^{[1\bar{1}\bar{1}]}$ whereas the response at J4 was expected to arise from $M_t^{[\bar{1}1\bar{1}]}$ and $M_t^{[\bar{1}11]}$ or vice versa as shown in Figure 3 (b). This was expected to be the underlying cause of sign reversal from J2 to J4. The overall behavior can be summed up as a waveform as shown in Figure 3 (b). It is noted that the non-local resistance increased as a function of temperature in MgO/p-Si sample whereas it decreased as a function of temperature in the Pt/MgO/p-Si sample. While both responses were expected to arise from topological phonons but this opposite behavior is still not well understood.

In conclusion, we presented experimental evidence of the spin superfluidity and second sound in the inhomogeneously strained p-Si samples. The stationary spin current from the topological electronic magnetism of phonons was expected to be the underlying cause of the spin superfluidity. The spin superfluidity was discovered using non-local resistance measurement. The non-local resistance measurement also showed a resonance behavior, which was attributed to the possible advent of second sound of topological phonons. The non-local resistance measurements showed a distribution similar to a stationary wave transport, which was attributed to the topological phonons.

**References**

[1]     P. C. Lou and S. Kumar, Journal of Physics: Condensed Matter **30**, 145801 (2018).

[2] L. Wang, S. Liu, X. Feng, C. Zhang, L. Zhu, J. Zhai, Y. Qin, and Z. L. Wang, Nature Nanotechnology **15**, 661 (2020).

[3] C. Schriever *et al.*, Advanced Optical Materials **3**, 129 (2015).

[4] M. Cazzanelli *et al.*, Nature Materials **11**, 148 (2011).

[5] P. C. Lou, A. Katailiha, R. G. Bhardwaj, W. P. Beyermann, D. M. Juraschek, and S. Kumar, Nano Letters **21**, 2939 (2021).

[6] Y. Ren, C. Xiao, D. Saparov, and Q. Niu, arXiv preprint arXiv:2103.05786 (2021).

[7] P. C. Lou, A. Katailiha, R. G. Bhardwaj, T. Bhowmick, W. P. Beyermann, R. K. Lake, and S. Kumar, Phys. Rev. B **101**, 094435 (2020).

[8] E. B. Sonin, Adv. Phys. **59**, 181 (2010).

[9] E. B. Sonin, J. Low Temp. Phys. **171**, 757 (2013).

[10] A. Beardo *et al.*, Science Advances **7**, eabg4677.